\documentclass[12pt]{iopart}
\usepackage{iopams,graphicx}  
\begin{document}
\title[Evaluation on asymptotic distribution of particle systems]{Evaluation on asymptotic distribution of particle systems expressed by probabilistic cellular automata}

\author{Kazushige Endo}

\address{Department of Pure and Applied Mathematics, Waseda University, 3-4-1, Okubo, Shinjuku-ku, Tokyo 169-8555, Japan}
\ead{k-endo@aoni.waseda.jp}
\begin{abstract}
We propose some conjectures for asymptotic distribution of probabilistic Burgers cellular automaton (PBCA) which is defined by a simple motion rule of particles including a probabilistic parameter. Asymptotic distribution of configurations converges to a unique steady state for PBCA. We assume some conjecture on the distribution and derive the asymptotic probability expressed by GKZ hypergeometric function. If we take a limit of space size to infinity, a relation between density and flux of particles for infinite space size can be evaluated. Moreover, we propose two extended systems of PBCA of which asymptotic behavior can be analyzed as PBCA. 
\end{abstract}
\noindent{\it keywords\/}: {cellular automaton, dynamical system, stochastic process, hypergeometric function}
\section{Introduction}
Cellular automata (CA) are dynamical systems with discrete time, discrete space and a finite set of state values. Their dynamics is generally determined by a simple rule which depends on values of neighboring space sites. Though this simple construction, they give fruitful mathematical properties and have been studied in various theoretical and applied research fields. For example, systems called `elementary cellular automata' (ECA) was classified according to their behavior of solutions and have been vastly studied by many researchers\cite{wolfram}.\par

There exist 256 independent rules for ECA. One of the rules referred by the rule number 184 is known as a non-trivial particle system. It is also called Burgers cellular automaton (BCA) since this can be derived from Burgers equation utilizing ultradiscretization method which was discovered in the field of integrable systems\cite{nishinari}. There exists a threshold of the density of particles and the asymptotic behavior of solution drastically changes between the regions of lower and higher density. Thus, a phase transition occurs for BCA at the threshold. Considering the dynamics of BCA as a transportation system of cars, this phase transition can be interpreted as a primitive model on occurrence of traffic jam.\par

We can introduce probabilistic parameters into the deterministic CA and they also have the vast theoretical and applied themes.  For example, the asymmetric simple exclusive process (ASEP) is a well-known standard statistical model for a simple stochastic particle system. It is a random walk model of multiple particles where each particle moves to neighboring sites with an excluded volume effect. Various exact evaluation for statistical results have been shown as for ASEP. Sasamoto et al. revealed exact relations between ASEP and the orthogonal polynomials\cite{sasamoto}.\par

  On the other hand, there exist many probabilistic CA giving the realistic dynamical systems as applications. For example, Nagel and Shreckenberg proposed a quite efficient dynamical model to investigate the physics of traffic jam and it is known as Nagel-Shreckenberg (NS) model\cite{schreckenberg}. They applied their model to the real freeway traffic and obtained a good accordance between the observed data and its theoretical estimation\cite{freeway}.\par
  The author and his co-workers analyzed the asymptotic behavior of probabilistic CA with 4 neighbors or with higher order of conserved quantities\cite{kuwabara,endo}. There exist some bilinear equations among probabilities of local patterns in the asymptotic solutions to the systems. Utilizing these equations, they derived a theoretical expression of relation between density and flux for the asymptotic behavior, which is called fundamental diagram (FD). \par

  In this paper, we focus on probabilistic extension of BCA (PBCA). It is partially equivalent to the `totally' ASEP (TASEP) obtained by restricting the motion of particles of ASEP to the only one direction\cite{derrida1,derrida2,kanai}. However, there is a difference between PBCA and TASEP about the updating way of particle positions.  Though one of particles to be updated is chosen every time step in TASEP, the `parallel-update' is used for PBCA, that is, motions of all particles from current time to the next time are determined simultaneously.\par

  We report a new analysis in order to understand asymptotic behavior of parallel-updated PBCA with the periodic boundary condition. We can consider PBCA as a one-dimensional random process and derive a transition matrix for the process. Assuming the random process is ergodic, we propose a conjecture about the asymptotic distribution of the system. Using the conjecture, we can derive the asymptotic probability of each configuration of particles in space sites and can derive FD from their expected values. Moreover, we give the expression of FD of PBCA by a hypergeometric function proposed by Gelfand, Kapranov and Zelevinsky.  It is called GKZ hypergeometric function and are obtained by extending the hypergeometric function of a single variable to multiple variables\cite{GKZ}. They obey the specific form of differential equations and their contiguous relation can be obtained in the form of matrix\cite{kakei}. The FD of PBCA with infinite space sites is calculated by utilizing the limit of the contiguous relations. This limiting case coincides with that of precedence research based on a specific ansatz\cite{schreckenberg}.  Furthermore, we propose two kinds of extension of PBCA and derive the asymptotic probability of configurations and FD using the similar conjecture of PBCA.

  Contents of this paper are as follows.  In section 2, we introduce definition and properties of PBCA, and present the new type of analysis for PBCA. In section 3, we propose two kinds of extensions of PBCA, and give the results based on the similar conjecture of PBCA. In section 4, we give the conclusion.  In the appendix A, we show calculation of PBCA in the limit of infinite space size utilizing GKZ hypergeometric function.
\section{Probabilistic Burgers cellular automaton}
\subsection{Definition of particle system}
Probabilistic Burgers cellular automaton (PBCA) is expressed by the following max-plus equation,
\begin{equation}
u_j^{n+1}=u_j^n +q_{j-1}^n-q_{j}^n,\qquad u_j^n \in \{0,1\},
\end{equation}
where
\begin{equation*}
q_j^{n}=\min(a_j^n,u_j^n,1-u_{j+1}^n),
\qquad
 a_j^n=
\cases{1 & (prob. $\alpha$) \\
0 & ($1-\alpha$)}.
\end{equation*}
The subscript $j$ is an integer site number and the superscript $n$ an integer time.  The real constant $\alpha$ satisfies $0\leq \alpha \leq 1$ and the value of probabilistic parameter $a_j^n$ is determined independently for every $(j,n)$.  We assume a periodic boundary condition for space sites with a period $L$, that is, $u_{j+L}^n=u_j^n$. From the above evolution equation, it is easily shown that
\begin{equation}
  \sum_{j=1}^{L}u_j^{n+1}=\sum_{j=1}^{L}u_j^{n}.
\end{equation}
Thus, the sum of all state values over $L$ is conserved for $n$ which is determined by the initial data. Let us consider that $u_j^n$ means the number of particle at site $j$ and time $n$.  Then a motion rule of particles for PBCA can be interpreted as follows.
\begin{center}
  \setlength\unitlength{1truecm}
  \begin{picture}(1,1)(0,0)
  \put(0,0){10}
  \put(0.08,0.3){\includegraphics[scale=0.015]{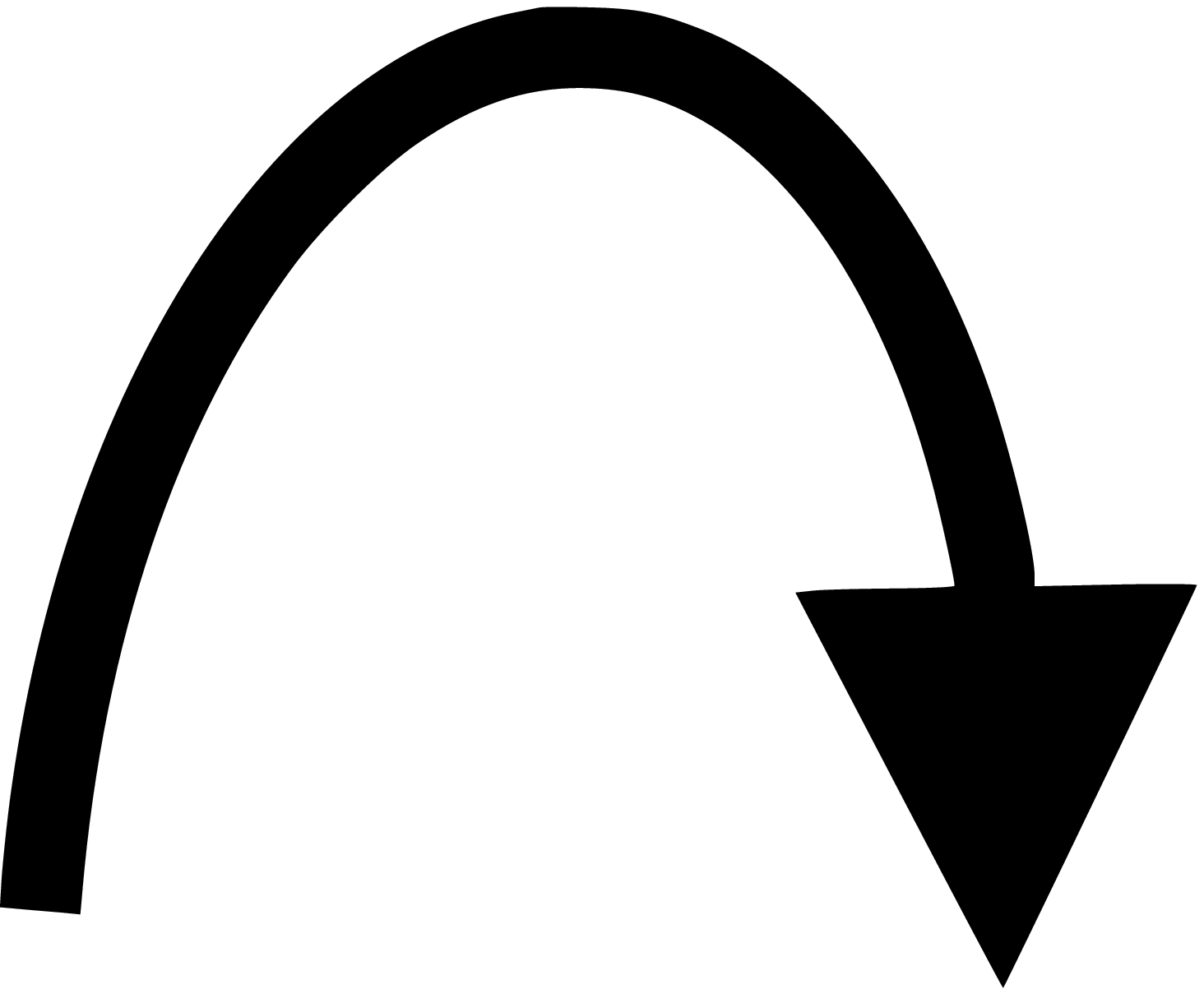}}
  \put(-0.4,0.6){prob. $\alpha$}
  \end{picture}
\end{center}
A particle at $j$th site moves to right with probability $\alpha$ only if a particle does not exist at ($j+1$)th site. Figure~\ref{fig:pbcate} shows an example of time evolution of PBCA.
\begin{figure}[htb]
\begin{center}
  \setlength\unitlength{1truecm}
  \begin{picture}(4,4.5)(1.5,0)
  \put(-0.27,4){\vector(1,0){2}}
  \put(2,3.9){$j$}
  \put(-0.27,4){\vector(0,-1){2}}
  \put(-0.35,1.63){$n$}
  \put(0,0){\includegraphics[width=70mm]{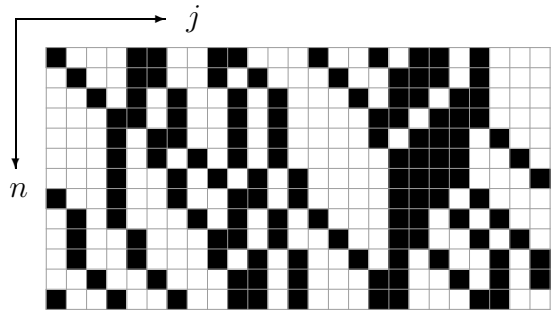}}
  \end{picture}
\end{center}
\caption{Example of time evolution of PBCA for $\alpha=0.5$.  Black squares $\blacksquare$ mean $u=1$ and white squares $\square$ $u=0$.}
\label{fig:pbcate}
\end{figure}
\subsection{Asymptotic distribution of PBCA}
\begin{figure}[!h]
\begin{tabular}{cc}
  \includegraphics[width=70mm]{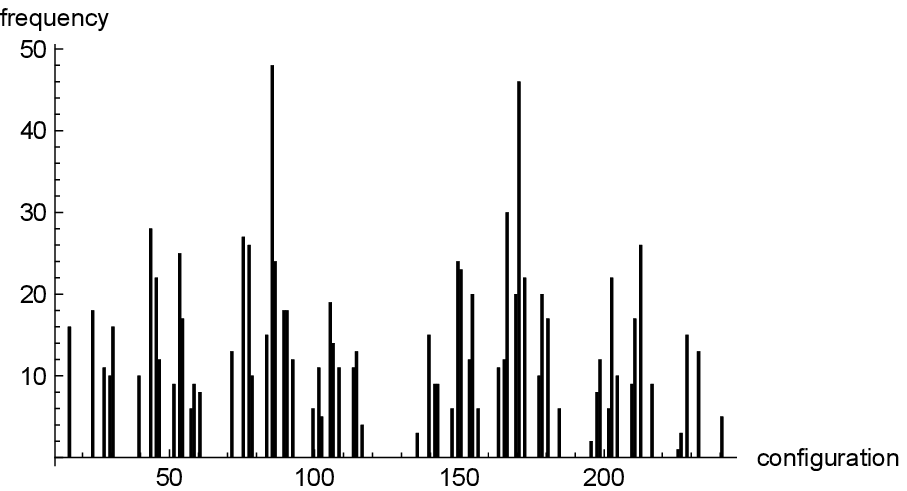}
&
  \includegraphics[width=70mm]{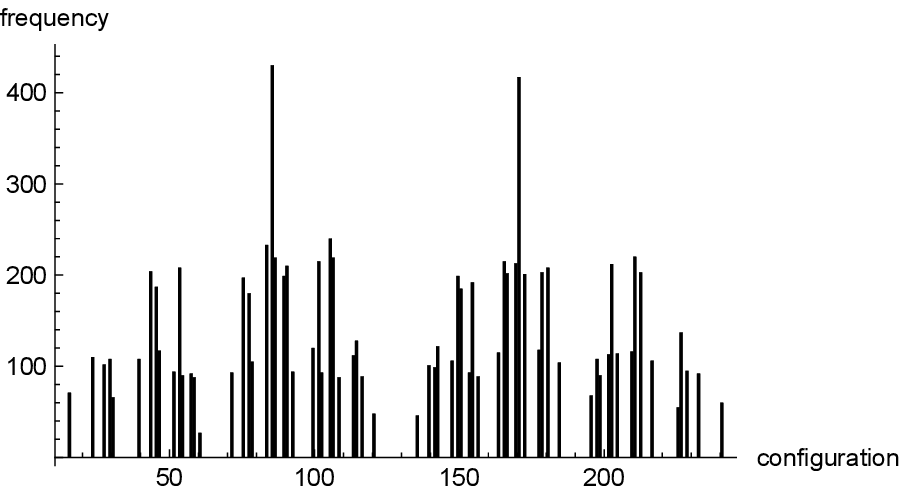}
\\
  (a) $0\le n\le1000$ & (b) $0\le n\le10000$
\medskip\\
  \includegraphics[width=70mm]{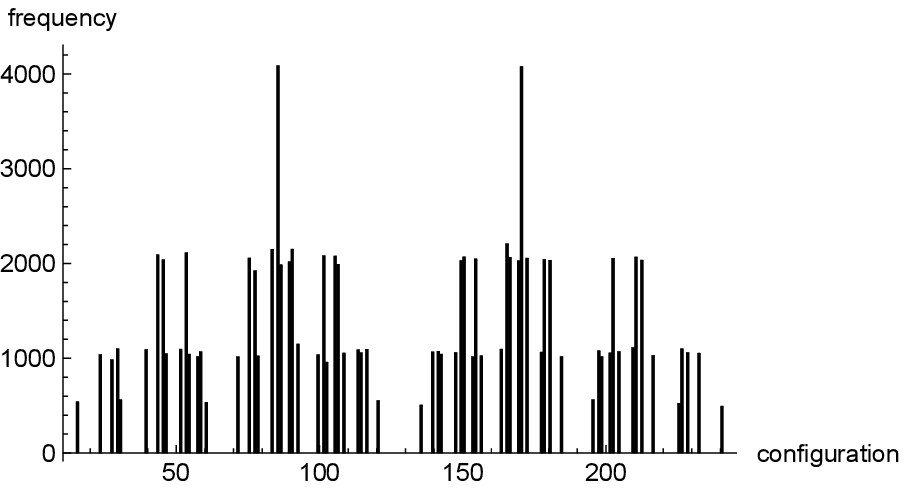}
&
  \includegraphics[width=70mm]{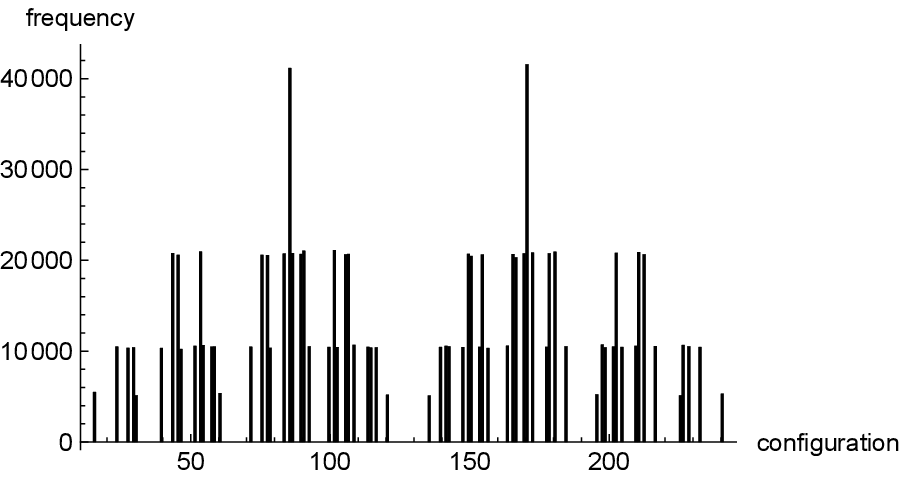}
\\
  (c) $0\le n\le100000$ & (b) $0\le n\le1000000$
\end{tabular}
\caption{Histograms of configurations for $L=8$, $m=4$ and $\alpha=0.5$.}  \label{fig:hist}
\end{figure}

Supposing that a set of values over all $L$ sites corresponds to a configuration of random process, PBCA is a one-dimensional random process on ${}_L \mathrm{C}_m$ configurations if $m=\sum_{j=1}^Lu_j^n$, that is, if $m$ particles move through the sites. Figure~\ref{fig:hist} (a) shows a histogram of all configurations from $n=0$ to $n=1000$ obtained by a numerical computation. Figures~\ref{fig:hist} (b), (c) and (d) show those from n=0 to n=10000, 100000 and 1000000, respectively. These figures suggest that the distribution of configurations converges after large enough time steps. Moreover, heights of bins can be considered to be categorized by some classes of heights after enough time steps.
We made various numerical calculations on the histogram and an approximately unique steady state is always obtained. Therefore, we can assume that PBCA is ergodic. To ensure our assumption and to explain the relation  among heights of classes, we show below a few concrete and exact results for small $L$ and $m$.\par
  For $L=4$ and $m=2$, a set of all configurations which is denoted by $\Omega$ is
\begin{equation*}
\Omega=\{0011,\ 0110,\ 1100,\ 1001,\ 0101,\ 1010\},
\end{equation*}
and transition probabilities are obtained as
\begin{eqnarray*}
1010 \quad &\rightarrow& \quad 0101 \quad \hbox{with probability} \quad \alpha^2 \\
1010 \quad &\rightarrow& \quad 1001 \quad \hbox{with probability} \quad \alpha(1-\alpha) \\
1010 \quad &\rightarrow& \quad 0110 \quad \hbox{with probability} \quad \alpha(1-\alpha) \\
1010 \quad &\rightarrow& \quad 1010 \quad \hbox{with probability} \quad (1-\alpha)^2 \\
&\vdots&
\end{eqnarray*}
Then, a transition matrix $A=(a_{ij})$ is
\begin{equation*}
\begin{array}{c}
0011\\
0110\\
1100\\
1001\\
0101\\
1010
\end{array}
\left(
\begin{array}{cccccc}
1-\alpha & 0 & 0 & 0 & 0 & \alpha \\
0 & 1-\alpha & 0 & 0 & \alpha & 0 \\
0 & 0 & 1-\alpha & 0 & 0 & \alpha \\
0 & 0 & 0 & 1-\alpha & \alpha & 0 \\
\alpha(1-\alpha) & 0 & \alpha(1-\alpha) & 0 & (1-\alpha)^2 & \alpha^2 \\
0 & \alpha(1-\alpha) & 0 & \alpha(1-\alpha) & \alpha^2 & (1-\alpha)^2 \\
\end{array}
\right),
\end{equation*}
where the element $a_{ij}$ is a transition probability from the $i$th configuration to the $j$th configuration.  The characteristic equation for the matrix has a simple root 1, and the absolute value of other roots is always less than 1. Then, time evolution of the random process determined by this matrix has a unique steady state, and an eigenvector of the matrix for eigenvalue 1 becomes the asymptotic distribution of configurations. The corresponding eigenvector  is
\begin{equation*}
(1-\alpha,\ 1-\alpha,\ 1-\alpha,\ 1-\alpha,\ 1,\ 1)
\end{equation*}
and components of this eigenvector give the ratios of heights of histogram.\par
  Let us consider another example of $L=6$ and $m=3$.  All elements of a set of configurations $\Omega$ are
\begin{equation*}
\begin{array}{l}
000111,\ 001110,\ 011100,\ 111000, \ 110001,\ 100011,\ 001011,\ 010110,\\
101100,\ 011001,\ 110010,\ 100101,\ 010011,\ 100110,\ 001101,\ 011010,\\
110100,\ 101001,\ 010101,\ 101010.
\end{array}
\end{equation*}
The transition matrix is
\begin{equation*}
\begin{array}{c}
000111\\
001110\\
011100\\
\vdots \\
101010
\end{array}
\left(
\begin{array}{ccccc}
1 - \alpha & 0 & 0 & \ldots & 0 \\
0 & 1 - \alpha & 0 & \ldots & 0 \\
0 & 0 & 1 - \alpha & \ldots & 0 \\
\vdots & \vdots & \vdots & \ddots & \vdots \\
0 & 0 & 0 & \ldots & (1-\alpha)^3 \\
\end{array}
\right),
\end{equation*}
and its eigenvector for the eigenvalue 1 as a simple root of characteristic equation is
\begin{equation*}
\big(\overbrace{(1-\alpha)^2,\ldots(1-\alpha)^2}^{\hbox{\small 6 components}},\overbrace{1-\alpha,\ldots 1-\alpha}^{\hbox{\small 12 components}}, 1, 1\big).
\end{equation*}
\par
  We propose the following conjecture from exact results obtained for small $L$ and numerical results for large $L$ and $m$.  To give the conjecture, let us introduce some notations. Define $\Omega$ by a set of all configurations for $L$ and $m$, $x=x_1  x_2\cdots x_L$ ($x_i \in \{ 0,1 \}$) by any configuration of $\Omega$, $\#s_1 s_2\cdots s_k(x)$ ($1 \leq k \leq L$) by the number of patterns $s_1 s_2\cdots s_k$ included in $x$ considering the periodic boundary condition, and $p(x)$ by the probability of $x$ in the steady state.
\begin{description}
\item[Conjecture:]
For any $x\in\Omega$, we have
\begin{equation*}
p(x)=\frac{C}{(1-\alpha)^{\#10(x)}},
\end{equation*}
where $C$ is a normalization constant satisfying $\sum_{x\in\Omega} p(x)=1$.
\end{description}
Considering the case of $L=4$, and $m=2$ as an example, we have 
\begin{eqnarray*}
\#10=1&:&p(0011)=p(0110)=p(1100)=p(1001)=\frac{C}{1-\alpha}, \\
\#10=2&:&p(0101)=p(1010)=\frac{C}{(1-\alpha)^2}.
\end{eqnarray*}
Since $p(0011)+p(0110)+\cdots+p(1010)=1$, we have $C=(1-\alpha)^2/(4(1-\alpha)+2)$ and
\begin{equation*}
p(0011)=\frac{1-\alpha}{4(1-\alpha)+2}, \qquad p(0101)=\frac{1}{4(1-\alpha)+2}.
\end{equation*}
Considering another case of $L=11$, and $m=6$, we have
\begin{eqnarray*}
\#10=1&:&p(11111100000)=p(01111110000)=\cdots=\frac{C}{1-\alpha}, \\
\#10=2&:&p(11110110000)=p(10011100011)=\cdots=\frac{C}{(1-\alpha)^2}, \\
\#10=3&:&p(11101101000)=p(11011001100)=\cdots=\frac{C}{(1-\alpha)^3}, \\
\cdots.
\end{eqnarray*}
\par
  By the above conjecture, we can categorize the probability by $\#10$. Then the probability of a configuration $x$ with $\#10(x)$ obtained in the limit $n\to\infty$ for the space size $L$ and the number of particles $m$ is
\begin{equation*}
  p(x)=\frac{(\frac{1}{1-\alpha})^{\#10(x)}}{\sum_{k=1}^m N_{L,m}(k)(\frac{1}{1-\alpha})^{k}} \qquad (0<\alpha<1),
\end{equation*}
where $N_{L,m}(k)$ is the number of configurations with $\#10=k$ defined by
\begin{equation*}
N_{L,m}(k)=\frac{L(m-1)!(L-m-1)! }{(m-k)!(k-1)!k! (L-m-k)!}.
\end{equation*}
\subsection{Fundamental diagram of PBCA}
\begin{figure}[htbp]
\begin{center}
\includegraphics[width=70mm]{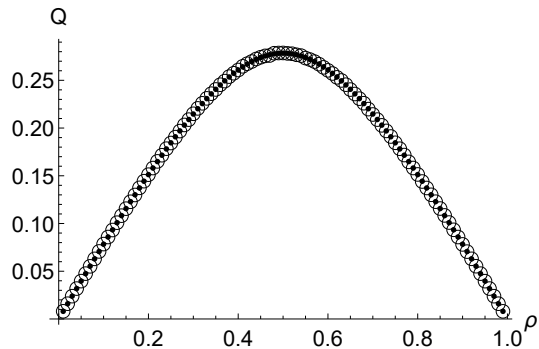}
\end{center}
\caption{Example of FD. Small black circles ($\bullet$) are obtained by (\ref{fd_pbca}) for $L=100$ and $\alpha=0.8$, and white circles ($\bigcirc$) are obtained numerically for the same $L$ and $\alpha$ averaged from $n=0$ to 50000.}
  \label{fig:pbca}
\end{figure}

Fundamental diagram (FD) of general particle systems is a diagram which is the relation between density and flux of particles averaged over all sites in the limit $n\to\infty$. Since PBCA is also a particle system, we can derive its FD. The density $\rho$ is defined by $m/L$ where $L$ and $m$ are the number of sites and of particles respectively.  Define $Q_{L,\alpha}(m)$ by the expected values of flux of the steady state where $\alpha$ is the hopping probability of particle. Then $Q_{L,\alpha}(m)$ of PBCA is an expected value of $\alpha\#10/L$ since only particles next to empty site can move with probability $\alpha$. Thus, we have
\begin{equation}
\label{fd_pbca}
Q_{L,\alpha}(m)=\frac{\alpha }{L}\frac{\sum_{k=1}^{m} k\,N_{L,m}(k)(\frac{1}{1-\alpha})^k}{\sum_{k=1}^{m} N_{L,m}(k)(\frac{1}{1-\alpha})^k} \qquad (0<m<L).
\end{equation}
Figure~\ref{fig:pbca} shows FD obtained by (\ref{fd_pbca}) and that by the numerical calculation.  The former is shown by small black circles ($\bullet$) and the latter by white circles ($\bigcirc$).  Their good coincidence can be observed from this figure.\par
Utilizing GKZ hypergeometric function, we can evaluate $Q_{L,\alpha}(m)$ in the limit of $L\to\infty$ preserving $m=\rho L$ as
\begin{equation}
Q=\lim_{L\to \infty\atop m=\rho L} Q_{L,\alpha}(m)=\frac{1-\sqrt{1-4\alpha\rho(1-\rho)}}{2}.
\end{equation}
We show the proof about this derivation in the appendix A\cite{kakei}.
\section{Extended systems and their properties}
In this section, we introduce two systems which are extensions of PBCA. We evaluate their asymptotic probability of configurations and derive their FD as we have shown in the previous section. Note that we also assume the ergodicity for both systems.
\subsection{Extension to 4 neighbors}
Let us consider the probabilistic CA expressed by a max-plus equation,
\begin{equation*}
u_j^{n+1}=u_j^n +q_{j-1}^n-q_{j}^n,
\end{equation*}
where
\begin{equation}
q_j^n=\max(\min(u_{j}^n,1-u_{j+1}^n,u_{j+2}^n,b_j^n),\min(u_{j}^n,1-u_{j+1}^n,1-u_{j+2}^n,a_j^n)).
\end{equation}
The probabilistic parameters $a_j^n$ and $b_j^n$ are defined by
\begin{equation*}
a_j^n=
\cases{
1 & (prob. $\alpha$) \\
0 & ($1-\alpha$)
},
\qquad b_j^n=
\cases{
1 & (prob. $\beta$) \\
0 & ($1-\beta$)
}.
\end{equation*}
Since this equation is also in the conservation form, the number of particles ($=\sum_{j} u_j^n$) is preserved for time evolution. From the max-plus expression of flow $q_j^n$, we can easily show the motion rule of particles denoted by 1 as follows.
\begin{center}
  \setlength\unitlength{1truecm}
  \begin{picture}(2,0.8)(0,0)
  \put(0,0){100}
  \put(0.08,0.3){\includegraphics[scale=0.015]{arrow7.eps}}
  \put(-0.3,0.6){prob. $\alpha$}
  \put(1.8,0){101}
  \put(1.88,0.3){\includegraphics[scale=0.015]{arrow7.eps}}
  \put(1.5,0.6){prob. $\beta$}
  \end{picture}
\end{center}
Note that any particle in the configuration other than $100$ and $101$ can not move. We call this system EPBCA1. In the case of $\alpha=\beta$, any particle in the patterns 10 moves with probability $\alpha$. Thus, PBCA is included in EPBCA1 as a special case.  Figure \ref{fig:epbca1} shows an example of time evolution of the system.\par
\begin{figure}[htb]
\begin{center}
  \setlength\unitlength{1truecm}
  \begin{picture}(4,4.5)(1.5,0)
  \put(-0.45,4){\vector(1,0){2}}
  \put(1.75,3.9){$j$}
  \put(-0.46,4){\vector(0,-1){2}}
  \put(-0.56,1.58){$n$}
  \put(0,0){\includegraphics[width=70mm]{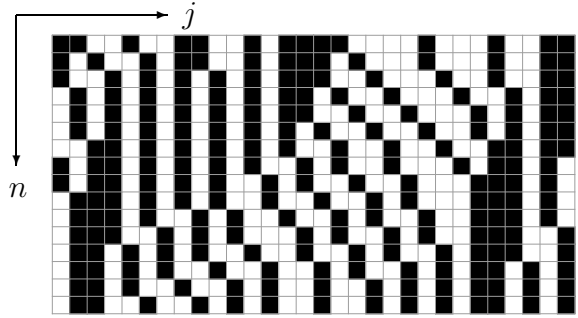}}
  \end{picture}
\end{center}
\caption{Example of time evolution of EPBCA1 for $\alpha=0.8$ and $\beta=0.1$.  Black squares $\blacksquare$ mean $u=1$ and white squares $\square$ $u=0$.}
\label{fig:epbca1}
\end{figure}
  We can obtain an exact form of transition matrix and of eigenvector for eigenvalue 1 for small values of the space size $L$ and the number of particles $m$. Then, we give a conjecture for asymptotic distribution of EPBCA1 from those results.  Let us introduce two examples suggesting our conjecture.\par
  Suppose a set of configurations $\Omega$ for $L=8$ and $m=4$. Size of $\Omega$ is 70. In order to make the expression of transition matrix compact, we define classes of configurations up to cyclic rotation and divide $\Omega$ into the equivalence classes. The number of the equivalence classes is 10, and their representative elements are as follows.
\begin{eqnarray*}
& 00001111,\ 00010111,\ 00011011,\ 00011101,\ 00100111, \\
& 00101011,\ 00101101,\ 00110011,\ 00110101,\ 01010101.
\end{eqnarray*}
Thus, a transition matrix derived by transition probabilities from each representative element to the others is
\begin{equation*}
\begin{array}{c}
00001111\\
00010111\\
00011011\\
\vdots \\
01010101
\end{array}
\left(
\begin{array}{ccccc}
1 - \alpha & 0 & 0 & \ldots & 0 \\
(1-\alpha)\beta & (1 - \alpha)(1-\beta) & 0 & \ldots & 0 \\
0 & (1-\alpha)\beta & (1 - \alpha)(1-\beta) & \ldots & 0 \\
\vdots & \vdots & \vdots & \ddots & \vdots \\
0 & 0 & 0 & \ldots & (1-\beta)^4+\beta^4 \\
\end{array}
\right),
\end{equation*}
and an eigenvector for the eigenvalue 1 is
\begin{eqnarray*}
& \Big(
\frac{4(1-\alpha)^2\beta^3(1-\beta)}{\alpha^3},\ 
\frac{4(1-\alpha)\beta^2(1-\beta)}{\alpha^2},\ 
\frac{4(1-\alpha)\beta^2(1-\beta)}{\alpha^2},\ \\
&\qquad \frac{4(1-\alpha)\beta^2(1-\beta)}{\alpha^2},\ 
\frac{4\beta^2(1-\beta)^2}{\alpha^2},\ 
\frac{4\beta(1-\beta)}{\alpha},\ 
\frac{4\beta(1-\beta)}{\alpha},\ \\
&\qquad \frac{2\beta^2(1-\beta)^2}{\alpha^2},\ 
\frac{4\beta(1-\beta)}{\alpha},\ 
1\Big).
\end{eqnarray*}
\par
  Another example is the case of $L=9$ and $m=3$. The size of $\Omega$ is 84. We define classes of configurations up to cyclic rotation and divide $\Omega$ into the equivalence classes. The number of the equivalence classes is 10, and their representative elements are
\begin{eqnarray*}
&000000111,\ 000001011,\ 000001101,\ 000010011,\ 000010101, \\
&000011001,\ 000100011,\ 000100101,\ 000101001,\ 001001001.
\end{eqnarray*}
Thus, a transition matrix is
\begin{equation*}
\begin{array}{c}
000000111\\
000001011\\
000001101\\
\vdots \\
001001001
\end{array}
\left(
\begin{array}{ccccc}
1 - \alpha & 0 & \alpha & \ldots & 0 \\
(1-\alpha)\beta & (1 - \alpha)(1-\beta) & \alpha\beta & \ldots & 0 \\
0 & (1-\alpha)\beta & (1 - \alpha)(1-\beta) & \ldots & 0 \\
\vdots & \vdots & \vdots & \ddots & \vdots \\
0 & 0 & 0 & \ldots & (1-\alpha)^3+\alpha^3 \\
\end{array}
\right),
\end{equation*}
and an eigenvector for the eigenvalue 1 is
\begin{eqnarray*}
&\Big(
\frac{3(1-\alpha)^4\beta^2}{\alpha^2(1-\beta)^2},\ 
\frac{3(1-\alpha)^3\beta}{\alpha(1-\beta)^2},\ 
\frac{3(1-\alpha)^3\beta}{\alpha(1-\beta)^2},\ 
\frac{3(1-\alpha)^2\beta}{\alpha(1-\beta)},\ 
\frac{3(1-\alpha)^2}{(1-\beta)^2},\ 
\\
&\qquad\qquad\frac{3(1-\alpha)^2\beta}{\alpha(1-\beta)},\ 
\frac{3(1-\alpha)^2\beta}{\alpha(1-\beta)},\ 
\frac{3(1-\alpha)}{1-\beta},\ 
\frac{3(1-\alpha)}{1-\beta},\ 
1\Big).
\end{eqnarray*}
Examining other concrete examples for small $L$, we propose a conjecture for the asymptotic distribution of configurations as follows.
\begin{description}
\item[Conjecture:]
The probability of any configuration $x\in\Omega$ in the steady state is given by
\begin{equation}
p(x)=C\,\left(\frac{\alpha(1-\beta)}{(1-\alpha)^2 \beta}\right)^{\#100(x)} \left(\frac{\alpha}{(1-\alpha) \beta}\right)^{\#101(x)},
\end{equation}
where $C$ is a normalization constant satisfying $\sum_{x\in\Omega} p(x)=1$.
\end{description}
\par
\begin{figure}[htbp]
\begin{center}
\includegraphics[width=70mm]{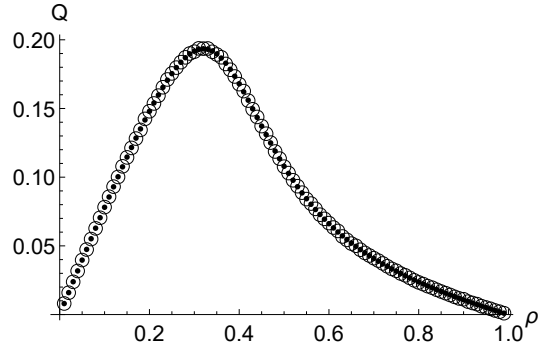}
\end{center}
\caption{Example of FD. Small black circles ($\bullet$) are obtained by (\ref{flux_epbca1}) for $L=100$, $\alpha=0.8$ and $\beta=0.1$, and white circles ($\bigcirc$) are obtained numerically for the same $L$, $\alpha$ and $\beta$ averaged from $n=0$ to 10000.}
  \label{fig:epbca1 fd}
\end{figure}

By the above conjecture, the probability of $x$ in the steady state of the space size $L$ and the number of particles $m$ is
\begin{equation*}
p(x)=
\frac{
(\frac{\alpha(1-\beta)}{(1-\alpha)^2\beta})^{\#100(x)}
(\frac{\alpha}{(1-\alpha)\beta})^{\#101(x)}
}
{
\sum\limits_{1\leq k_1+k_2\leq m\atop0\le L-m-2k_1-k_2}
N(k_1,k_2)(\frac{\alpha(1-\beta)}{(1-\alpha)^2\beta})^{k_1}
(\frac{\alpha}{(1-\alpha)\beta})^{k_2}
},
\end{equation*}
where
\begin{equation*}
N(k_1,k_2)=
\frac{L(L-m-k_1-k_2-1)!(m-1)!}
{k_1!k_2!(L-m-2k_1-k_2)!(k_1-1)!(m-k_1-k_2)!}.
\end{equation*}
Since the mean flux $Q_{L,\alpha,\beta}(m)$ is an expected value of $(\alpha \#100 + \beta \#101)/L$, it becomes
\begin{equation}
\label{flux_epbca1}
Q_{L,\alpha,\beta}(m)
=\frac{1}{L}\frac{
\sum\limits_{1\leq k_1+k_2\leq m\atop 0\le L-m-2k_1-k_2}
(\alpha k_1+\beta k_2)
N(k_1,k_2)(\frac{\alpha(1-\beta)}{(1-\alpha)^2\beta})^{k_1}
(\frac{\alpha}{(1-\alpha)\beta})^{k_2}
}
{\sum\limits_{1\leq k_1+k_2\leq m\atop 0\le L-m-2k_1-k_2}
N(k_1,k_2)(\frac{\alpha(1-\beta)}{(1-\alpha)^2\beta})^{k_1}
(\frac{\alpha}{(1-\alpha)\beta})^{k_2}
}.
\end{equation}
Figure~\ref{fig:epbca1 fd} shows FD obtained by (\ref{flux_epbca1})and that by the numerical calculation.  The former is shown by small black circles ($\bullet$) and the latter by white circles ($\bigcirc$).  Their good coincidence can be observed from this figure.
\subsection{Extension to 2 kinds of particles}
\newcommand{\0}{{\tt 0}}
\newcommand{\A}{{\tt A}}
\newcommand{\B}{{\tt B}}
Let us consider a probabilistic CA defined by the following system of max-plus equations.
\begin{eqnarray*} 
&&u_j^{n+1}=u_j^n+q_{j-1}^n-q_{j}^n, \qquad q_j^n=\min(a_j^n, u_{j-1}^n,1-u_j^n, 1-v_j^n), \\
&&v_j^{n+1}=v_j^n+r_{j-1}^n-r_{j}^n, \qquad r_j^n=\min(b_j^n, v_{j-1}^n,1-v_j^n, 1-u_j^n).
\end{eqnarray*}
Assume $u_j^n$, $v_j^n \in \{0,1\}$ and probabilistic parameters $a_j^n$ and $b_j^n$ are defined by
\begin{equation*}
a_j^n=
\cases{
1 & (prob. $\alpha$) \\
0 & ($1-\alpha$)
},\qquad
b_j^n=
\cases{
1 & (prob. $\beta$) \\
0 & ($1-\beta$)
}.
\end{equation*}
The above evolution rule can be interpreted as a probabilistic system of two kinds of particles considering that $u_j^n$ and $v_j^n$ are the number of particles of kind A and B respectively at site $j$ and time $n$. Moreover, let us assume these two kinds of particles do not exist at the same site and at the same time, that is, $(u_j^n,v_j^n)\ne(1,1)$. If we set this condition for the initial data, it is always satisfied through time evolution following to the above evolution rule. Below we express the configuration of particles $(u,v)=(0,0)$, $(1,0)$ and $(0,1)$ by the symbols $\0$, $\A$ and $\B$, respectively. Then the motion rule of particles is given by the following.
\begin{center}
  \setlength\unitlength{1truecm}
  \begin{picture}(3,1)(0,0)
  \put(0,0){\A\0}
  \put(0.1,0.3){\includegraphics[scale=0.015]{arrow7.eps}}
  \put(-0.2,0.6){prob. $\alpha$}
  \put(2.5,0){\B\0}
  \put(2.6,0.3){\includegraphics[scale=0.015]{arrow7.eps}}
  \put(2.26,0.6){prob. $\beta$}
  \end{picture}
\end{center} 
We call the above system EPBCA2. If $v_j^n\equiv 0$, the motion rule for $u_j^n$ of EPBCA2 reduces to that of PBCA and that for $v_j^n$ trivial.  Thus, PBCA is included in this system as a special case.\par
\begin{figure}[htb]
\begin{center}
  \setlength\unitlength{1truecm}
  \begin{picture}(4,4.5)(1.5,0)
  \put(-0.5,4){\vector(1,0){2}}
  \put(1.8,3.9){$j$}
  \put(-0.5,4){\vector(0,-1){2}}
  \put(-0.59,1.58){$n$}
  \put(0,0){\includegraphics[width=70mm]{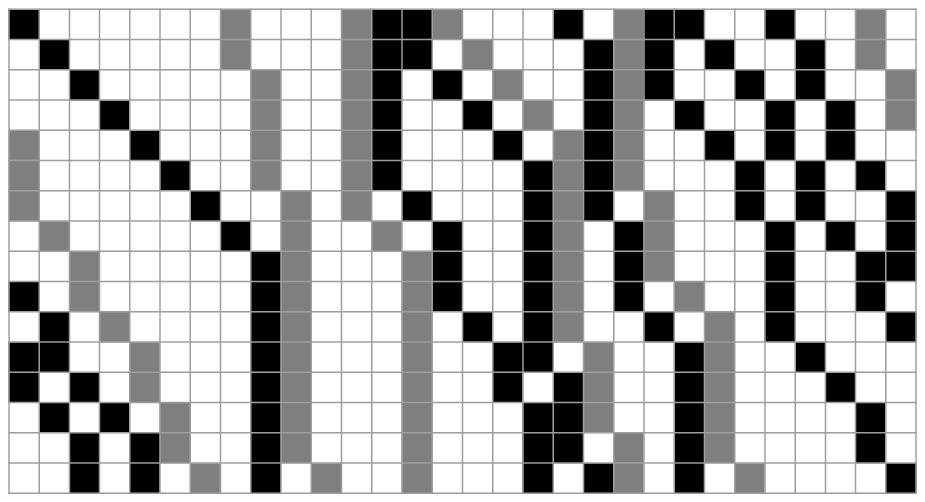}}
  \end{picture}
\end{center}
\caption{Example of time evolution of EPBCA2 for $\alpha=0.4$ and $\beta=0.8$. White, grey and black squares express $(u,v)=(0,0)$ (no particle), $(1,0)$ (only particle A exists) and $(0,1)$ (only particle B exists), respectively.}
\label{fig:epbca2}
\end{figure}
Figure~\ref{fig:epbca2} shows an example of time evolution of EPBCA2. Any particle does not pass over the other particles following to the evolution rule. Let us define a `sequence' by an array of particles from a given configuration preserving their relative positions. For example, the sequence in a configuration $\0\A\A\0\0\B\B\0$ is $\A\A\B\B$ and this sequence may change into the sequences $\B\A\A\B$, $\B\B\A\A$, and $\A\B\B\A$ along time evolution. However, the sequence $\A\A\B\B$ can not change into $\A\B\A\B$ since particles do not pass over one another. Similarly, the sequence $\A\B\A\B$ may change into $\B\A\B\A$ but not either $\A\A\B\B$, $\B\A\A\B$, $\B\B\A\A$ or $\A\B\B\A$. If we consider a transition matrix of configurations for a given initial data, the configurations are restricted to those obtained from the initial data. Therefore, if a set of configurations $\Omega$ include a configuration $x=x_1x_2\ldots x_L$, we restrict $\Omega$ to be constructed from any combinations of $\0$'s, $\A$'s and $\B$'s preserving their numbers and the order of sequence included in $x$ up to cyclic rotation. Below we give two examples of transition matrices and their eigenvectors.\par
  First example is $\Omega$ including a configuration $\A\A\B\A\A\B\0\0$. The size of $\Omega$ is 84. In order to make the expression of transition matrix compact, we define classes of configurations up to cyclic rotation and divide $\Omega$ into the equivalence classes. The number of equivalence classes is 12, and their representative elements are as follows.
\begin{eqnarray*}
&\0\0\A\A\B\A\A\B,\ \0\A\0\A\B\A\A\B,\ \0\A\A\0\B\A\A\B,\ \0\A\A\B\0\A\A\B,\ \0\A\A\B\A\0\A\B,\ \0\A\A\B\A\A\0\B, \\
&\A\0\0\A\B\A\A\B,\ \A\0\A\0\B\A\A\B,\ \A\0\A\B\A\0\A\B,\ \A\0\A\B\A\A\0\B,\ \A\A\0\0\B\A\A\B,\ \A\A\0\B\A\A\0\B.
\end{eqnarray*}
The transition matrix derived by transition probabilities from each representative element to the others is
\begin{equation*}
\begin{array}{c}
\0\0\A\A\B\A\A\B\\
\0\A\0\A\B\A\A\B\\
\0\A\A\0\B\A\A\B\\
\vdots \\
\A\A\0\B\A\A\0\B
\end{array}
\left(
\begin{array}{ccccc}
1-\beta & 0 & 0 & \ldots & 0 \\
\alpha(1-\beta) & (1-\alpha)(1-\beta) & 0 & \ldots & 0 \\
0 & \alpha(1-\beta) & (1-\alpha)(1-\beta) & \ldots & (1-\alpha)\beta \\
\vdots & \vdots & \vdots & \ddots & \vdots \\
0 & 0 & 0 & \ldots & (1-\alpha)^2 \\
\end{array}
\right),
\end{equation*}
and its eigenvector for the eigenvalue 1 is
\begin{eqnarray*}
&\Big(
\frac{8\alpha^2 (1-\beta)}{(1-\alpha)^2 \beta^2},\ 
\frac{8\alpha}{(1-\alpha)^2 \beta},\ 
\frac{8\alpha}{(1-\alpha)^2 \beta},\ 
\frac{4\alpha^2}{(1-\alpha)^2 \beta^2},\ 
\frac{8\alpha}{(1-\alpha)^2 \beta},\ 
\frac{8\alpha}{(1-\alpha)^2 \beta}, \\
& \frac{8}{1-\alpha},\ 
\frac{8}{(1-\alpha)^2},\ 
\frac{4}{(1-\alpha)^2},\ 
\frac{8}{(1-\alpha)^2},\ 
\frac{8}{1-\alpha},\ 
\frac{4}{(1-\alpha)^2} \Big).
\end{eqnarray*}
\par
  Another example is $\Omega$ including a configuration $\A\A\B\A\0\0\0$. The size of $\Omega$ is 140, the number of equivalence classes is 20, and their representative elements are
\begin{eqnarray*}
&&\0\0\0\A\A\B\A,\ \0\0\A\0\A\B\A,\ \0\0\A\A\0\B\A,\ \0\0\A\A\B\0\A,\ \0\A\0\0\A\B\A,\ \0\A\0\A\0\B\A, \\
&&\0\A\0\A\B\0\A,\ \0\A\A\0\0\B\A,\ \0\A\A\0\B\0\A,\ \0\A\A\B\0\0\A,\ \A\0\0\0\A\B\A,\ \A\0\0\A\0\B\A, \\
&&\A\0\0\A\B\0\A,\ \A\0\A\0\0\B\A,\ \A\0\A\0\B\0\A,\ \A\0\A\B\0\0\A,\ \A\A\0\0\0\B\A,\ \A\A\0\0\B\0\A, \\
&&\A\A\0\B\0\0\A,\ \A\A\B\0\0\0\A.
\end{eqnarray*}
The transition matrix is
\begin{equation*}
\begin{array}{c}
\0\0\0\A\A\B\A\\
\0\0\A\0\A\B\A\\
\0\0\A\A\0\B\A\\
\vdots \\
\A\A\B\0\0\0\A
\end{array}
\left(
\begin{array}{ccccc}
1-\alpha & 0 & 0 & \ldots & 0 \\
\alpha(1-\alpha) & (1-\alpha)^2 & 0 & \ldots & 0 \\
0 & \alpha(1-\alpha) & (1-\alpha)^2 & \ldots & 0 \\
\vdots & \vdots & \vdots & \ddots & \vdots \\
0 & 0 & 0 & \ldots & 1-\beta \\
\end{array}
\right),
\end{equation*}
and the eigenvector for the eigenvalue 1 is
\begin{eqnarray*}
&\Big(
\frac{1}{1-\alpha},
\frac{1}{(1-\alpha)^2},
\frac{1}{(1-\alpha)^2},
\frac{\alpha}{(1-\alpha)^2 \beta},
\frac{1}{(1-\alpha)^2},
\frac{1}{(1-\alpha)^3},
\frac{\alpha}{(1-\alpha)^3 \beta}, \\
&\frac{1}{(1-\alpha)^2},
\frac{\alpha}{(1-\alpha)^3 \beta},
\frac{\alpha^2 (1-\beta)}{(1-\alpha)^3 \beta^2},
\frac{1}{1-\alpha},
\frac{1}{(1-\alpha)^2},
\frac{\alpha}{(1-\alpha)^2 \beta},
\frac{1}{(1-\alpha)^2},\\
&\frac{\alpha}{(1-\alpha)^3 \beta},
\frac{\alpha^2 (1-\beta)}{(1-\alpha)^3 \beta^2},
\frac{1}{1-\alpha},
\frac{\alpha}{(1-\alpha)^2 \beta},
\frac{\alpha^2 (1-\beta)}{(1-\alpha)^3 \beta^2},
\frac{\alpha^3 (1-\beta)^2}{(1-\alpha)^3 \beta^3}
\Big).
\end{eqnarray*}
Examining other concrete examples for small $L$, we propose a conjecture for the asymptotic distribution of configurations. Before giving the conjecture, we set the notations.  Assume a space size is $L$ and the numbers of particle A and B are $m_\A$ and $m_\B$ respectively. If we choose a certain configuration, a set $\Omega$ of configurations is defined by all realizable configurations evolving from it. Let $x=x_1 x_2\ldots x_L$ be one of configurations in $\Omega$.  Define $k_\A$ and $k_\B$ by the number of local patterns $\A\0$ and $\B\0$ included in $x$ respectively. Define $n_\A$ by the number of all \0's in the local patterns $\A\0\ldots\0\B$ and $\A\0\ldots\0\A$, and $n_\B$ by $\B\0\ldots\0\A$ and $\B\0\ldots\0\B$. For example, a configuration
\begin{equation*}
\B\A\A\A\0\0\A\0\B\0\B\B\0\0\A\0\0\0,
\end{equation*}
is given, then $k_\A=5$, $k_\B=4$, $n_\A=6$ and $n_\B=3$. Note that $m_\A+m_\B+n_\A+n_\B=L$. Using these notations, we give the following conjecture.
\begin{description}
\item[Conjecture:]
The probability of configuration $x$ in the steady state is
\begin{equation*}
p(x) = C\,\frac{(1-\beta)^{n_\B-k_\B}}{(1-\alpha)^{k_\A+n_\B}}
\Big(\frac{\alpha}{\beta}\Big)^{n_\B},
\end{equation*}
where $C$ is a normalization constant satisfying $\sum_{x\in\Omega} p(x)=1$.
\end{description}
From the above conjecture, the probability of any configuration $x$ for $k_\A$, $k_\B$, $n_\A$ and $n_\B$ in the steady state is given by
\begin{equation}
p(x)=
\frac{
  \frac{(1-\beta)^{n_\B-k_\B}}{(1-\alpha)^{k_\A+n_\B}}
  (\frac{\alpha}{\beta})^{n_\B}
}
{
\sum\limits_{k_\A \leq \min(n_\A, m_\A),\atop
             {k_\B \leq \min(n_\B, m_\B),\atop
             n_\A+n_\B=L-m_\A-m_\B}}
N(k_\A,k_\B,n_\A,n_\B)
\frac{(1-\beta)^{n_\B-k_\B}}{(1-\alpha)^{k_\A+n_\B}}
(\frac{\alpha}{\beta})^{n_\B}
},
\end{equation}
where $N(k_\A, k_\B, n_\A, n_\B)$ denotes the number of all configurations included in $\Omega$ with $k_\A$, $k_\B$, $n_\A$ and $n_\B$ and is calculated as
\begin{eqnarray*}
&& N(k_\A,k_\B,n_\A,n_\B)= \\
&& \displaystyle\frac{D\cdot m_\A!(n_\A-1)!m_\B!(n_\B-1)!}
{(m_\A-k_\A)!k_\A!(n_\A-k_\A)!(k_\A-1)!(m_\B-k_\B)!k_\B!(n_\B-k_\B)!(k_\B-1)!},
\end{eqnarray*}
where $D$ is a constant dependent on $\Omega$.\par
\begin{figure}[htbp]
\begin{center}
\begin{tabular}{cc}
  \includegraphics[width=70mm]{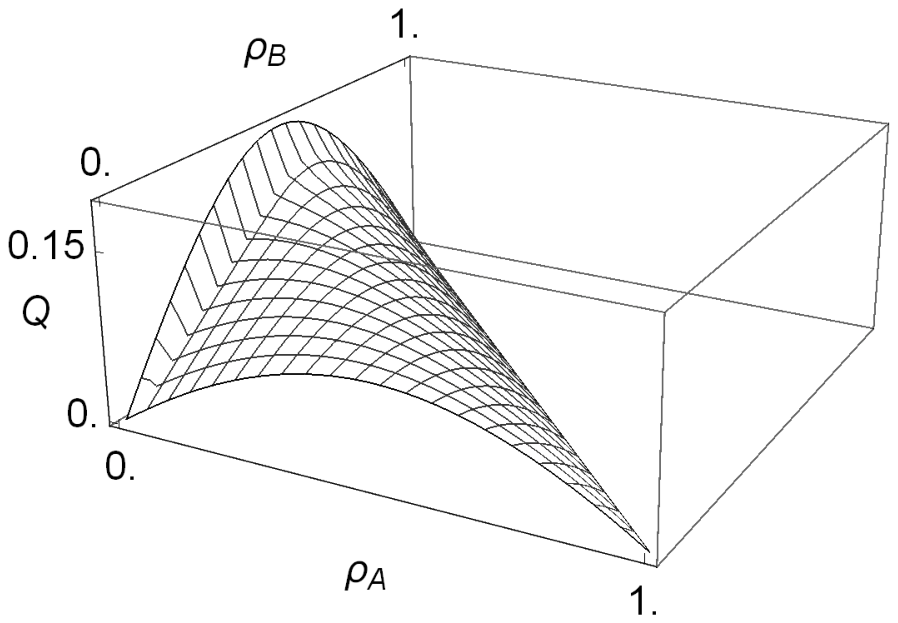}
&
  \includegraphics[width=70mm]{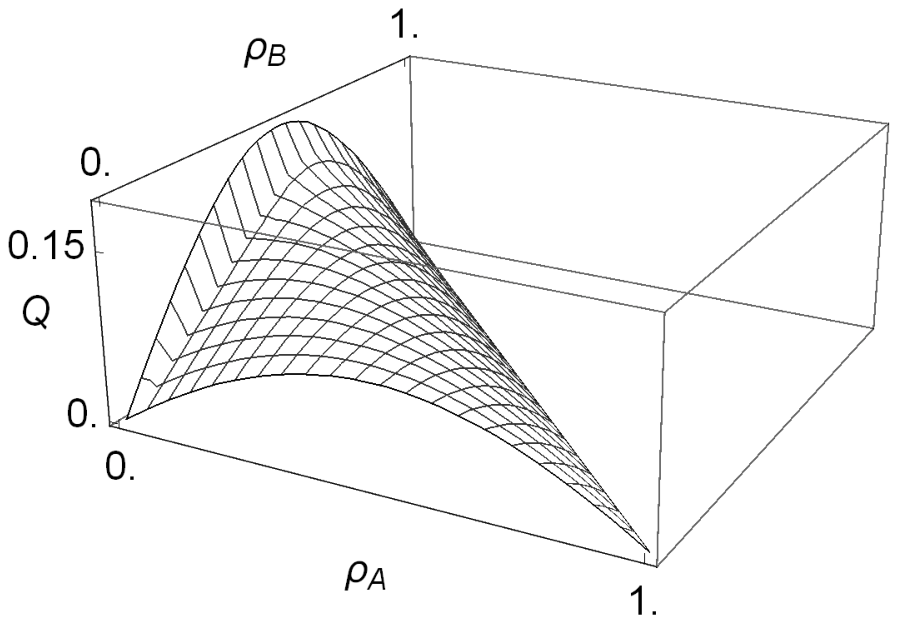} \\
(a) & (b)
\end{tabular}
\end{center}
\caption{Example of FD. (a) Theoretical result by (\ref{flux_epbca2}) for $L=30$, $\alpha=0.3$ and $\beta=0.6$, (b) Numerical result averaged from $n=0$ to $n=100000$ with the same parameters of (a).}
\label{fig:epbca2 fd}
\end{figure}
\begin{figure}[htbp]
\begin{center}
\includegraphics[width=70mm]{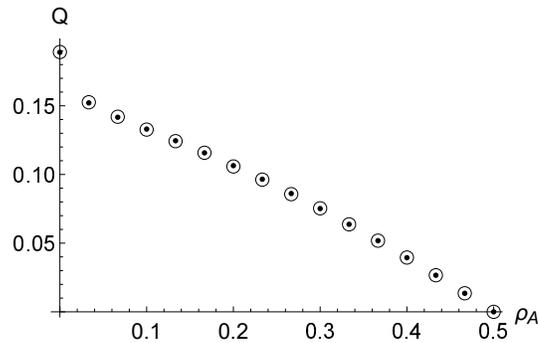}
\end{center}
\caption{Example of FD for $\rho_\B=0.5$. Small black circles ($\bullet$) are obtained by (\ref{flux_epbca2}) for $L=30$, $\alpha=0.3$ and $\beta=0.6$, and white circles ($\bigcirc$) are obtained numerically for the same $L$, $\alpha$ and $\beta$ averaged from $n=0$ to 100000.}
  \label{fig:epbca2 fd2}
\end{figure}

  Considering the motion rule of particle A and B, the flux in the steady state is 
\begin{equation} \label{flux_epbca2}
Q_{L,\alpha,\beta}(m_\A,m_\B)=
\frac{
\sum\limits_{k_\A \leq \min(n_\A, m_\A),\atop
             {k_\B \leq \min(n_\B, m_\B),\atop
             n_\A+n_\B=L-m_\A-m_\B}}
\frac{\alpha k_\A +\beta k_\B}{L}
N(k_\A, k_\B, n_\A, n_\B)
\frac{(1-\beta)^{n_\B-k_\B}}{(1-\alpha)^{k_\A+n_\B}}
(\frac{\alpha}{\beta})^{n_\B}
}
{
\sum\limits_{k_\A \leq \min(n_\A, m_\A),\atop
             {k_\B \leq \min(n_\B, m_\B),\atop
             n_\A+n_\B=L-m_\A-m_\B}}
N(k_\A,k_\B,n_\A,n_\B)
\frac{(1-\beta)^{n_\B-k_\B}}{(1-\alpha)^{k_\A+n_\B}}
(\frac{\alpha}{\beta})^{n_\B}
}.
\end{equation} 
Figure~\ref{fig:epbca2 fd} (a) shows FD calculated by (\ref{flux_epbca2}) and figure~\ref{fig:epbca2 fd} (b) shows the numerical result. In these figures, $\rho_A$ and $\rho_B$ are densities of particles $A$ and $B$ respectively.
Figure~\ref{fig:epbca2 fd2} shows a comparison of FD's of figure~\ref{fig:epbca2 fd} for $\rho_\B=0.5$. Small black circles ($\bullet$) are obtained by (\ref{flux_epbca2}) and white circles ($\bigcirc$) by the numerical calculation.  Their good coincidence can be observed from this figure.
\section{Concluding remarks}
  We gave some conjectures for the asymptotic distribution of PBCA and its extended systems assuming that these systems are ergodic. In the conjectures, the following points are the most important.
\begin{itemize}
\item Probability of any configuration in the steady state depends only on the number of some specific local pattern included in the configuration. For example, it depends only on $\#10$ for PBCA.
\item Since the probabilities do not depend on location of the patterns in the configuration, they are equal each other if the numbers of patterns included in configurations are also.  For example, the probabilities of configurations are the same if their $\#10$ are the same in the case of PBCA.
\end{itemize}
Based on the conjecture, we derived the probability of configuration in the steady state and FD of the system for arbitrary size of $L$.\par
  On the other hand, FD of PBCA for $L\to \infty$ is reported in the previous research as
\begin{equation*}
  Q=\frac{1-\sqrt{1-4\alpha \rho(1-\rho)}}{2},
\end{equation*}
using some ansatz on relations of probabilities of local patterns\cite{schreckenberg}.  Utilizing GKZ hypergeometric function, we evaluate the limit of FD of PBCA (\ref{fd_pbca}) and confirm that the diagram coincides with the above result.\par
  Furthermore, since FD of the extended systems is also expressed by some kinds of GKZ hypergeometric function, we expect that FD in the limit of infinite size can be derived similarly as PBCA. In particular, for EPBCA2, for the initial condition of sequence $\ldots\A\B\A\B\A\B\A\B\ldots$ and a probabilistic parameter $\alpha =1$, motion rule of particles in the steady state becomes as follows.
\begin{center}
  \setlength\unitlength{1truecm}
  \begin{picture}(2.5,1)(0,0)
  \put(0,0){\A\0\B}
  \put(0.12,0.3){\includegraphics[scale=0.015]{arrow7.eps}}
  \put(-0.287,0.6){prob. 1}
  \put(2,0){\B\0}
  \put(2.12,0.3){\includegraphics[scale=0.015]{arrow7.eps}}
  \put(1.72,0.6){prob. $\beta$}
  \end{picture}
\end{center}
This motion rule is the same as that of the L-R system obtained by ultradiscrete Cole-Hopf transformation of SECA84 with a quadratic conserved quantity\cite{endo}. The FD of L-R system in the limit of infinite space size is evaluated and derived in a simple form depending on the density of particles. Therefore, FD of EPBCA2 in the limit of $L\to\infty$ includes FD of SPCA84 as a special case.
\appendix
\section{Limit of FD of PBCA utilizing GKZ hypergeometric function}
\subsection{Definition of GKZ hypergeometric function and its properties}  \label{subsec:GKZ}
In this section,  we introduce definition of GKZ hypergeometric function and its properties. First, let us define $B_{k}^{\gamma}(x)$ by
\begin{equation*}
B_{k}^{\gamma}(x)=\frac{x^{\gamma+k}}{\Gamma (\gamma+k+1)}.
\end{equation*}
It satisfies the following contiguous relations.
\begin{eqnarray*}
&& \frac{d}{dx}B_{k}^{\gamma}(x)=B_{k-1}^{\gamma}(x)=B_{k}^{\gamma-1}(x), \\
&& x B_{k}^{\gamma}(x)=(\gamma+k+1)B_{k+1}^{\gamma}(x)=(\gamma+k+1)B_{k}^{\gamma+1}(x).
\end{eqnarray*}
For an $m\times n$ matrix $A=(a_{ij})$ and an $n$-dimensional vector $\vec{\beta}$, general solution to
\begin{equation*}
A\vec{\gamma}=\vec{\beta},
\end{equation*}
is expressed by
\begin{equation*}
\vec{\gamma}=\vec{\gamma_0}+\vec{k}, \quad \vec{k}\in \ker A,
\end{equation*}
where $\vec{\gamma_0}$ is a special solution to the equation.
Using $A$, $\vec{\beta}$, $\vec{\gamma_0}$ and $\vec{k}$, GKZ hypergeometric function $\varphi(\vec{\beta};\vec{x})$ is defined by
\begin{equation}
\varphi(\vec{\beta};\vec{x})=\sum_{\vec{k}\in \ker A \cap \mathbb{Z}^n} \prod_{j=1}^{n} B_{k_j}^{\gamma_{0,j}}(x_j),
\end{equation}
where $\vec{x}$ is an $n$-dimensional vector and $\gamma_{0,j}$, $k_j$ and $x_j$ are $j$th element of $\vec{\gamma_0}$ , $\vec{k}$ and $\vec{x}$ respectively. Utilizing contiguous relations, the following two properties for GKZ hypergeometric function are derived.\par
First, for a given vector
\begin{equation*}
\vec{b}=
\left(
\begin{array}{c}
b_1\\
b_2\\
\vdots \\
b_n
\end{array}
\right)
\in \ker A \cap \mathbb{Z}^n,
\end{equation*}
$J_{+}(\vec{b})$ and $J_{-}(\vec{b})$ are defined by subsets of $\{0,1,\cdots,n \}$ as
\begin{equation*}
J_{+}(\vec{b})=\{j|b_j > 0 \} ,\qquad J_{-}(\vec{b})= \{j|b_j < 0\}.
\end{equation*}
Then, we obtain
\begin{eqnarray*}
\prod_{j \in  J_{-}(\vec{b})} (\frac{\partial}{\partial x_j})^{-b_j}\varphi 
&=& \sum_{\vec{k}\in \ker A \cap \mathbb{Z}^n} \prod_{j \in  J_{-}(\vec{b})} (\frac{\partial}{\partial x_j})^{-b_j} \prod_{j=1}^n B_{k_j}^{\gamma_{0,j}}(x_j) \\
&=& \sum_{\vec{k}\in \ker A \cap \mathbb{Z}^n} \prod_{j \in  J_{+}(\vec{b})} B_{k_j}^{\gamma_{0,j}}(x_j) \prod_{j \in  J_{-}(\vec{b})}  B_{k_j+b_j}^{\gamma_{0,j}}(x_j) \\
&=& \sum_{\vec{k}\in \ker A \cap \mathbb{Z}^n} \prod_{j \in  J_{+}(\vec{b})} (\frac{\partial}{\partial x_j})^{b_j} \prod_{j=1}^n B_{k_j}^{\gamma_{0,j}}(x_j) \\
&=&\prod_{j \in  J_{+}(\vec{b})} (\frac{\partial}{\partial x_j})^{b_j}\varphi.
\end{eqnarray*}
Therefore, $\varphi$ satisfies a differential equation
\begin{equation}
\label{de0}
\left\{ \prod_{j \in  J_{+}(\vec{b})} (\frac{\partial}{\partial x_j})^{b_j}-\prod_{j \in  J_{-}(\vec{b})} (\frac{\partial}{\partial x_j})^{-b_j} \right\} \varphi=0.
\end{equation}\par
Second, for an operator $\theta_j=x_j\frac{\partial}{\partial x_j}$ and $\vec{\theta}$ defined by
\begin{equation*}
\vec{\theta}=
\left(
\begin{array}{c}
\theta_1\\
\theta_2\\
\vdots \\
\theta_n
\end{array}
\right),
\end{equation*}
$\varphi$ satisfies
\begin{equation*}
A\vec{\theta}\varphi(\vec{\beta};\vec{x})
=A(\vec{\gamma}+\vec{k})\varphi=\vec{\beta}\varphi.
\end{equation*}
Using $\alpha \in \mathbb{C}^{\times}$ and the $i$th row vector $\vec{a_i}=(a_{i1},\cdots,a_{in})$ of $A$, define $\alpha^{D(\vec{a_i})}$ by
\begin{equation*}
\alpha^{D(\vec{a_i})}= \hbox{diag}(\alpha^{a_{i1}},\cdots,\alpha^{a_{in}}).
\end{equation*}
Since
\begin{equation*}
B_{k_j}^{\gamma_j} (\alpha^{a_{ij}} x_j) = \alpha^{a_{ij} (\gamma_j + k_j)} B_{k_j}^{\gamma_j}(x_j),
\end{equation*}
$B_{k_j}^{\gamma_j} (x_j)$ satisfies 
\begin{equation*}
\prod_{j=1}^{n}B_{k_j}^{\gamma_j} (\alpha^{a_{ij}} x_j) = \alpha^{\sum_{j=1}^{n}a_{ij} (\gamma_j + k_j)} \prod_{j=1}^{n} B_{k_j}^{\gamma_j}(x_j) = \alpha^{\beta_i} \prod_{j=1}^{n} B_{k_j}^{\gamma_j}(x_j).
\end{equation*}
Thus,
\begin{equation}
\label{id}
\varphi(\vec{\beta};\alpha^{D(\vec{a_i})} \vec{x}) = \alpha^{\beta_i}\varphi(\vec{\beta};\vec{x}),
\end{equation}
is obtained.
\subsection{FD of PBCA expressed by GKZ hypergeometric function}
In this subsection, we express FD of PBCA by GKZ hypergeometric function. The FD of PBCA (\ref{fd_pbca}) for infinite space size is
\begin{equation*}
Q=\lim_{L\to \infty \atop m=\rho L}Q_{L,\alpha}(m)=\lim_{L\to\infty\atop m=\rho L}\frac{\alpha }{L}\frac{\sum_{k=1}^{m} k (\frac{1}{1-\alpha})^{k-1} N_{L,m}(k)}{\sum_{k=1}^{m} (\frac{1}{1-\alpha})^{k-1} N_{L,m}(k)},
\end{equation*}
where the density $\rho$ is constant and 
\begin{equation*}
N_{L,m}(k)=\frac{(m-1)!L(L-m-1)! }{(m-k)!(k-1)!k! (L-m-k)!}.
\end{equation*}
To evaluate the above limit, we choose the following matrix A and vector $\vec{\beta_1}$ in the definition of GKZ hypergeometric function in \ref{subsec:GKZ}.
\begin{equation*}
A = \left(
\begin{array}{cccc}
      0 & -1 & 1 & 0 \\
      1 & 1 & 1 & 1 \\
      1 & 0 & 1 & 0
\end{array}
\right),
\qquad
\vec{\beta_1}=
\left(
\begin{array}{c}
1\\
L-1\\
m
\end{array}
\right).
\end{equation*}
The general solution to $A \vec{\gamma}=\vec{\beta_1}$ is
\begin{equation*}
\vec{\gamma}=\vec{\gamma_0}+\vec{k}\qquad (\vec{k}\in \ker A \cap \mathbb{Z}^4 ),
\end{equation*}
where
\begin{equation}
\label{gamma}
\vec{\gamma_0}=
\left(
\begin{array}{c}
\gamma_{0,1}\\
\gamma_{0,2}\\
\gamma_{0,3}\\
\gamma_{0,4}\\
\end{array}
\right)
=
\left(
\begin{array}{c}
m\\
-1\\
0 \\
L-m
\end{array}
\right),
\qquad
\vec{k}=
\left(
\begin{array}{c}
k_{1}\\
k_{2}\\
k_{3}\\
k_{4}\\
\end{array}
\right)
=
k
\left(
\begin{array}{c}
-1\\
1\\
1 \\
-1
\end{array}
\right)
\quad(k\in \mathbb{Z}).
\end{equation}
Introducing a new notation $\varphi_{\beta_1, \beta_2, \beta_3}(x_1, x_2, x_3, x_4)$ defined by $\varphi(\vec{k}, \vec{x})$ for $\vec{\beta}=(\beta_1, \beta_2, \beta_3)$ and $\vec{x}=(x_1, x_2, x_3, x_4)$ and $\lambda$ by $1/(1-\alpha)$, we obtain
\begin{eqnarray*}
\varphi_{1,L-1,m}(1,\lambda,1,1)&=&\sum_{\vec{k}\in \ker A \cap \mathbb{Z}^4} B_{k_1}^{\gamma_{0,1}}(1) B_{k_2}^{\gamma_{0,2}}(\lambda) B_{k_3}^{\gamma_{0,3}}(1) B_{k_4}^{\gamma_{0,4}}(1) \\
&=&\sum_{k=1}^{m} \frac{\lambda^{k-1}}{(m-k)! (k-1)! k! (L-m-k)!}.
\end{eqnarray*}
Considering $\varphi_{\beta_1, \beta_2, \beta_3}(1,\lambda,1,1)$ as the function on $\lambda$, let us introduce a notation $F_{\beta_1, \beta_2, \beta_3}(\lambda)=\varphi_{\beta_1, \beta_2, \beta_3}(1,\lambda,1,1)$.\par
On the other hand, the general solution to
$A \vec{\gamma'}=\vec{\beta_2}$ for
\begin{equation*}
\vec{\beta_2}=
\left(
\begin{array}{c}
0\\
L-2\\
m-1
\end{array}
\right),
\end{equation*} is
\begin{equation*}
\vec{\gamma'}=\vec{\gamma'_0}+\vec{k}\qquad (\vec{k}\in \ker A \cap \mathbb{Z}^4 ),
\end{equation*}
where
\begin{equation}
\label{gamma_dash}
\vec{\gamma'_0}=
\left(
\begin{array}{c}
\gamma'_{0,1}\\
\gamma'_{0,2}\\
\gamma'_{0,3}\\
\gamma'_{0,4}\\
\end{array}
\right)
=
\left(
\begin{array}{c}
m\\
-1\\
-1 \\
L-m
\end{array}
\right).
\end{equation}
Therefore,
\begin{eqnarray*}
\varphi_{0,L-2,m-1}(1,\lambda,1,1)&=&\sum_{\vec{k}\in \ker A \cap \mathbb{Z}^4} B_{k_1}^{\gamma'_{0,1}}(1) B_{k_2}^{\gamma'_{0,2}}(\lambda) B_{k_3}^{\gamma'_{0,3}}(1) B_{k_4}^{\gamma'_{0,4}}(1) \\
&=&\sum_{k=1}^{m} \frac{(\frac{1}{1-\alpha})^{k-1}}{(m-k)! (k-1)! (k-1)! (L-m-k)!} \\
&=&F_{0,L-2,m-1}(\lambda),
\end{eqnarray*}
is obtained. Thus, we have
\begin{equation}
Q=\lim_{L\to\infty\atop m=\rho L}\frac{\alpha}{L}\frac{F_{0,L-2,m-1}(\lambda)}{F_{1,L-1,m}(\lambda)}.
\end{equation}
\subsection{Differential equation on $F_{1,L-1,m}$}
In this section, we derive a differential equation on $F_{1,L-1,m}$. Since the relations
\begin{eqnarray*}
\varphi_{1,L-1,m}(x_1, \frac{x_2}{a}, a x_3, x_4)&=&a \varphi_{1,L-1,m}(x_1,x_2,x_3,x_4), \\
\varphi_{1,L-1,m}(b x_1, b x_2, b x_3, b x_4)&=&b^{L-1}\varphi_{1,L-1,m}(x_1,x_2,x_3,x_4), \\
\varphi_{1,L-1,m}(c x_1, x_2, c x_3, x_4)&=&c^m \varphi_{1,L-1,m}(x_1,x_2,x_3,x_4),
\end{eqnarray*}
for $\varphi_{1,L-1,m}(x_1,x_2,x_3,x_4)$ and $a$, $b$ ,$c \in \mathbb{C}^{\times}$ are obtained from (\ref{id}), the relation
\begin{equation*}
\varphi_{1,L-1,m}(bc x_1,\frac{b}{a}x_2,abc x_3,b x_4)=ab^{L-1} c^{m} \varphi_{1,L-1,m}(x_1,x_2,x_3,x_4),
\end{equation*}
holds. If we assume
\begin{equation*}
(bc x_1,\frac{b}{a}x_2,abc x_3,b x_4)=(1,\lambda,1,1),
\end{equation*}
that is,
\begin{equation}
\label{abc}
a=\frac{x_1}{x_3},\quad b=\frac{1}{x_4},\quad c=\frac{x_4}{x_1},\quad \lambda=\frac{x_2 x_3}{x_1 x_4},
\end{equation}
the relation between $\varphi_{1,L-1,m}$ and $F_{1,L-1,m}$ is
\begin{equation}
\label{F}
\varphi_{1,L-1,m}(x_1,x_2,x_3,x_4)=x_1^{m-1} x_3 x_4^{L-m-1} F_{1,L-1,m}(\lambda).
\end{equation}
Moreover, we can derive
\begin{eqnarray*}
\frac{\partial^2}{\partial x_2 \partial x_3} \varphi_{1,L-1,m} &=&2x_1^{m-2} x_3 x_4^{L-m-2} F'_{1,L-1,m}(\lambda)+x_1^{m-3} x_2 x_3^2 x_4^{L-m-3} F''_{1,L-1,m}(\lambda), \\
\frac{\partial^2}{\partial x_1 \partial x_4} \varphi_{1,L-1,m} &=&(m-1)(L-m-1)x_1^{m-2} x_3 x_4^{L-m-2}F_{1,L-1,m}(\lambda) \nonumber \\
&-&(L-m-1)x_1^{m-3} x_2 x_3^2 x_4^{L-m-3} F'_{1,L-1,m}(\lambda) \nonumber \\
&-&(m-2)x_1^{m-3} x_2 x_3^2 x_4^{L-m-3} F'_{1,L-1,m}(\lambda) \nonumber \\
&+&x_1^{m-4} x_2^2 x_3^3 x_4^{L-m-4} F''_{1,L-1,m}(\lambda).
\end{eqnarray*}
Substituting these equations into the differential equation
\begin{equation*}
(\frac{\partial^2}{\partial x_2 \partial x_3} - \frac{\partial^2}{\partial x_1 \partial x_4})\varphi_{1,L-1,m}=0,
\end{equation*}
which is derived from (\ref{de0}), the following differential equation on $F_{1,L-1,m}$ is obtained.
\begin{eqnarray}
\label{de}
&\lambda (1-\lambda) F''_{1,L-1,m}(\lambda)+\{ (L-3)\lambda +2 \} F'_{1,L-1,m}(\lambda) \nonumber \\
&\qquad\qquad-(m-1)(L-m-1) F_{1,L-1,m}(\lambda)=0.
\end{eqnarray}
\subsection{Contiguous relation between $F_{1,L-1,m}$ and $F_{0,L-2,m-1}$}
From (\ref{gamma}) and (\ref{gamma_dash}), we have
\begin{equation*}
\vec{\gamma'}-\vec{\gamma}=
\left(
\begin{array}{c}
0\\
0\\
-1 \\
0
\end{array}
\right)
+k
\left(
\begin{array}{c}
-1\\
1\\
1 \\
-1
\end{array}
\right)
\qquad(k\in \mathbb{Z}).
\end{equation*}
Therefore,
\begin{eqnarray}
\varphi_{0,L-2,m-1}(x_1,x_2,x_3,x_4)&=&\sum_{\vec{k}\in \ker A \cap \mathbb{Z}^4} B_{k_1}^{\gamma'_{0,1}}(x_1) B_{k_2}^{\gamma'_{0,2}}(x_2) B_{k_3}^{\gamma'_{0,3}}(x_3) B_{k_4}^{\gamma'_{0,4}}(x_4) \nonumber \\
&=&\sum_{\vec{k}\in \ker A \cap \mathbb{Z}^4} B_{k_1}^{\gamma_{0,1}}(x_1) B_{k_2}^{\gamma_{0,2}}(x_2) B_{k_3}^{\gamma_{0,3}-1}(x_3) B_{k_4}^{\gamma_{0,4}}(x_4) \nonumber \\
&=&\frac{\partial}{\partial x_3} \varphi_{1,L-1,m}(x_1,x_2,x_3,x_4), \label{1}
\end{eqnarray}
is derived. From (\ref{id}), since the relations
\begin{eqnarray*}
\varphi_{0,L-2,m-1}( x_1,\frac{1}{a}x_2,a x_3, x_4)&=&\varphi_{0,L-2,m-1}(x_1,x_2,x_3,x_4), \\
\varphi_{0,L-2,m-1}(bx_1,bx_2,bx_3,bx_4)&=& b^{L-2}\varphi_{0,L-2,m-1}(x_1,x_2,x_3,x_4), \\
\varphi_{0,L-2,m-1}(c x_1,x_2,c x_3,x_4)&=&c^{m-1} \varphi_{0,L-2,m-1}(x_1,x_2,x_3,x_4),
\end{eqnarray*}
are obtained, the relation
\begin{equation*}
\varphi_{0,L-2,m-1}(bc x_1,\frac{b}{a}x_2,abc x_3,b x_4)= b^{L-2} c^{m-1} \varphi_{0,L-2,m-1}(x_1,x_2,x_3,x_4),
\end{equation*}
holds. Substituting (\ref{abc}) into this equation, we have
\begin{equation}
\label{2}
\varphi_{0,L-2,m-1}(x_1,x_2,x_3,x_4)=x_1^{m-1} x_4^{L-m-1} F_{0,L-2,m-1}(\lambda).
\end{equation}
Thus, we obtain
\begin{equation}
\label{3}
x_1^{m-1} x_4^{L-m-1} F_{0,L-2,m-1}(\lambda)=\frac{\partial}{\partial x_3} \varphi_{1,L-1,m}(x_1,x_2,x_3,x_4),
\end{equation}
from (\ref{1}) and (\ref{2}).\par
  On the other hand, derivative of (\ref{F}) with respect to $x_3$ is
\begin{eqnarray}
&&\frac{\partial}{\partial x_3} \varphi_{1,L-1,m}(x_1,x_2,x_3,x_4) \nonumber \\
&&\quad=x_1^{m-1}x_4^{L-m-1} F_{1,L-1,m}(\lambda)+x_1^{m-2} x_2 x_3 x_4^{L-m-2} F'_{1,L-1,m}(\lambda).
\label{4}
\end{eqnarray}
Therefore, from (\ref{3}) and (\ref{4}),
\begin{equation}
\label{neighbor}
F_{0,L-2,m-1}(\lambda)=F_{1,L-1,m}(\lambda)+\lambda F'_{1,L-1,m}(\lambda),
\end{equation}
is obtained. Derivative of (\ref{neighbor}) with respect to $\lambda$ is
\begin{equation}
\label{x}
F'_{0,L-2,m-1}(\lambda)= 2F'_{1,L-1,m}(\lambda) + \lambda F''_{1,L-1,m}(\lambda).
\end{equation}
Substituting (\ref{de}) into (\ref{x}) gives
\begin{equation}
\label{neighbordel}
F'_{0,L-2,m-1}(\lambda)=\frac{(m-1)(L-m-1)}{1-\lambda} F_{1,L-1,m}(\lambda)-\frac{(L-1)\lambda}{1-\lambda}F'_{1,L-1,m}(\lambda).
\end{equation}
From (\ref{neighbor}) and (\ref{neighbordel}), contiguous relation between $F_{1,L-1,m}$ and $F_{0,L-2,m-1}$ is 
\begin{equation}
\label{mat}
\left(
\begin{array}{c}
F_{0,L-2,m-1}(\lambda)\\
F_{0,L-2,m-1}'(\lambda)
\end{array}
\right)
=
\left(
\begin{array}{cc}
      1 & \lambda \\
      \frac{(m-1)(L-m-1)}{1-\lambda} & \frac{(L-1)\lambda}{1-\lambda}
\end{array}
\right)
\left(
\begin{array}{c}
F_{1,L-1,m}(\lambda)\\
F_{1,L-1,m}'(\lambda)
\end{array}
\right).
\end{equation}
\subsection{Limit of contiguous relation}
From the contiguous relation (\ref{mat}), we have
\begin{equation}
\label{mat2}
\left(
\begin{array}{c}
F_{0,L-2,m-1}(\lambda)\\
\frac{F'_{0,L-2,m-1}(\lambda)}{m}
\end{array}
\right)
=
\left(
\begin{array}{cc}
      1 & \lambda m \\
      \frac{(m-1)(L-m-1)}{(1-\lambda) m} & \frac{(L-1)\lambda}{1-\lambda}
\end{array}
\right)
\left(
\begin{array}{c}
F_{1,L-1,m}(\lambda)\\
\frac{F'_{1,L-1,m}(\lambda)}{m}
\end{array}
\right).
\end{equation}
Dividing both side of (\ref{mat2}) by $LF_{1,L-1,m}(\lambda)$,
\begin{equation}
\label{mateq}
\left(
\begin{array}{c}
\frac{F_{0,L-2,m-1}(\lambda)}{L F_{1,L-1,m}(\lambda)}\\
\frac{F'_{0,L-2,m-1}(\lambda)}{m L F_{1,L-1,m}(\lambda)}
\end{array}
\right)
=
\left(
\begin{array}{cc}
      \frac{1}{L} & \frac{\lambda m}{L} \\
      \frac{(m-1)(L-m-1)}{(1-\lambda) m L} & \frac{(L-1)\lambda}{(1-\lambda) L}
\end{array}
\right)
\left(
\begin{array}{c}
1\\
\frac{F'_{1,L-1,m}(\lambda)}{m F_{1,L-1,m}(\lambda)}
\end{array}
\right),
\end{equation}
is obtained. If we define
\begin{equation*}
g=\frac{F'_{1,L-1,m}(\lambda)}{F_{1,L-1,m}(\lambda)},
\end{equation*}
and substitute $g$ into (\ref{de}), we obtain
\begin{equation*}
\lambda(1-\lambda)(g'+g^2)+\{ (L-3)\lambda+2 \} g-(m-1)(L-m-1)=0.
\end{equation*}
Since $\rho=m/L$, we have
\begin{equation}
\label{eq_g}
\lambda(1-\lambda)(g'+g^2)+\{ (\frac{m}{\rho}-3)\lambda+2 \} g-(m-1)(\frac{m}{\rho}-m-1)=0.
\end{equation}
We can assume the following expansion of $g$ for $m=\infty$,
\begin{equation*}
g=g_1 m+g_0 + g_{-1}m^{-1} +g_{-2}m^{-2} +\cdots.
\end{equation*}
From the balance of $\mathcal{O}(m^2)$ terms of (\ref{eq_g}), we have
\begin{equation*}
\lambda(1-\lambda)g_1^2+\frac{\lambda}{\rho}g_1+1-\frac{1}{\rho}=0.
\end{equation*}
Solving this relation,
\begin{equation}
g_1=\frac{-\frac{\lambda}{\rho}-\sqrt{(\frac{\lambda}{\rho})^2-4\lambda(1-\lambda)(1-\frac{1}{\rho})}}{2\lambda(1-\lambda)},
\end{equation}
is obtained. Therefore, we can derive the limit of the first component of (\ref{mateq}) utilizing $g_1$ as
\begin{equation}
\lim_{L\to\infty\atop m=\rho L}\frac{\alpha }{L}\frac{F_{0,L-2,m-1}(\lambda)}{ F_{1,L-1,m}(\lambda)}=\alpha \rho \lambda g_1=\frac{1-\sqrt{1-4\alpha \rho(1-\rho)}}{2}.
\end{equation}
\section*{Acknowledgment}
We are grateful to Professor Saburo Kakei about the evaluation on the limit of FD of PBCA using GKZ hypergeometric function.
\section*{References}

\end{document}